\documentclass[aps,prl,twocolumn,tightenlines,amsmath,amssymb,superscriptaddress]{revtex4-2}

\usepackage{graphicx}
\usepackage{amsmath}
\usepackage{amssymb}
\usepackage{epstopdf}
\usepackage{color}
\usepackage{lipsum}
\usepackage{ulem}
\usepackage[dvipsnames]{xcolor}
\usepackage[UKenglish]{babel}
\usepackage[UKenglish]{isodate}
\usepackage[retainorgcmds]{IEEEtrantools}
\newcommand{\im}{\mathrm{i}}
\DeclareGraphicsExtensions{.eps}
\usepackage{braket}
\usepackage{color}
\usepackage{enumitem}
\usepackage{kbordermatrix}
\renewcommand{\kbldelim}{(}% Left delimiter
\renewcommand{\kbrdelim}{)}% Right delimiter

\begin{document}

\title{A {deterministic and efficient} source of frequency-polarization hyper-encoded photonic qubits}

\author{N. Coste$^{1,2}$, D. A. Fioretto$^{1,4}$, S. E. Thomas$^{1,3}$, S. C. Wein$^4$, H. Ollivier$^1$, I. Maillette de Buy Wenniger$^{1,3}$, A. Henry$^1$, N. Belabas$^1$, A. Harouri$^1$, A. Lemaitre$^1$, \\
I. Sagnes$^1$, N. Somaschi$^4$,  O. Krebs$^1$, L. Lanco$^{1,5,6}$ and P. Senellart}

\affiliation{Centre for Nanosciences and Nanotechnology, CNRS, Universite Paris-Saclay, UMR 9001,
10 Boulevard Thomas Gobert, 91120, Palaiseau, France\\
        $^2$School of Mathematical and Physical Sciences, University of Technology Sydney, Ultimo, New South Wales 2007, Australia\\
        $^3$Department of Physics, Imperial College London, London SW7 2BW, UK\\
        $^4$Quandela SAS, 10 Boulevard Thomas Gobert, 91120, Palaiseau, France\\
        $^5$Université Paris Cité, Centre for Nanoscience and Nanotechnology, F-91120 Palaiseau, France\\
        $^6$Institut Universitaire de France (IUF)}

	\begin{abstract}
		 
   {The frequency or color of photons is an attractive degree of freedom to encode and distribute the quantum information over long distances. However, the generation of frequency-encoded photonic qubits has so far relied on probabilistic non-linear single-photon sources and inefficient gates. Here, we demonstrate the deterministic generation of photonic qubits hyper-encoded in frequency and polarization based on a semiconductor quantum dot in a cavity. We exploit the double dipole structure of a neutral exciton and demonstrate the generation of any quantum superposition in amplitude and phase, controlled by the polarization of the pump laser pulse. The source generates frequency-polarization single-photon qubits at a rate of 4$\ $MHz corresponding to a generation probability at the first lens of $28 \pm 2\%$, with a photon number purity $>98\%$. The photons show an indistinguishability $>91\%$ for each dipole and $88\%$ for a balanced quantum superposition of both. The density matrix of the hyper-encoded photonic state is measured by  time-resolved polarization tomography, evidencing a fidelity to the target state of $94 \pm 8\%$ and concurrence of $77 \pm 2\%$, here limited by frequency overlap in our device. Our approach brings the advantages of quantum dot sources to the field of quantum information processing based on frequency encoding.}
		 \end{abstract}

\maketitle

{Photons are versatile carriers of quantum information for applications both in quantum communications~\cite{kimble_quantum_2008} or quantum computing~\cite{obrien_optical_2007}. A key advantage of photons is that they have many degrees of freedom to  flexibly encode the information tailored to specific use cases. Among them, frequency has recently attracted interest~\cite{lukens_frequency-encoded_2017}\cite{lu_frequency-bin_2023}, especially for quantum communication~\cite{bloch_frequency-coded_2007} since frequency is inherently stable over fiber links~\cite{williams_high-stability_2008}, and  is compatible with current demultiplexing technology  to build complex multiuser quantum networks~\cite{wengerowsky_entanglement-based_2018}. Because of those clear advantages, frequency qubits have been used to generate secure cryptographic keys, both using weak attenuated coherent pulses~\cite{bloch_frequency-coded_2007} and entangled pairs~\cite{cabrejo-ponce_high-dimensional_2023, henry_parallelization_2024}, and to connect distant quantum memories %, either 
via Bell state measurements~\cite{gao_quantum_2013} or Raman transitions~\cite{rosenblum_analysis_2017,chan_quantum_2022}}.

Significant progress has been made towards the generation and control of photonic qubits in the frequency domain. Sources of photonic frequency qubits have relied on probabilistic sources based on parametric downconversion and four-wave mixing that are intrinsically limited in efficiency by the presence of multi-photon components~\cite{kues_-chip_2017}. Deterministic single-qubit gates ~\cite{lu_electro-optic_2018} as well as two-qubit gates~\cite{lu_controlled-not_2019} have been demonstrated using nonlinear crystals, electro-optic modulators, and pulse shapers ~\cite{kobayashi_frequency-domain_2016}\cite{joshi_frequency-domain_2020} or electro-optic effects~\cite{lu_electro-optic_2018}, as well as Hong-Ou-Mandel interference~\cite{kobayashi_frequency-domain_2016}\cite{joshi_frequency-domain_2020}. These results lay the foundations for a complete architecture for quantum information processing in the frequency domain. However, the main challenges to real-world applications remain the losses introduced for performing those gates and the probabilistic nature of the sources. 

In this work, we take a meaningful step to address one of these  core challenges and demonstrate the deterministic and efficient generation of polarization and frequency hyper-encoded qubits. Our approach exploits semiconductor quantum dots{~\cite{senellart_high-performance_2017}} that are an excellent platform for the %deterministic
{on-demand} generation of pure and indistinguishable single photons \cite{tomm_bright_2021}\cite{maring_versatile_2024}\cite{ding_high-efficiency_2023}. We
{exploit} the energy levels of the neutral QD embedded into a micropillar cavity, which we address via an efficient, off-resonant excitation scheme assisted by the relaxation of Longitudinal Acoustical (LA) phonons~\cite{thomas_bright_2021}. In addition to {the superposition in the frequency domain}, the two %photonic
{single photon modes} are also orthogonal in polarization% (hyper-encoded)
, a feature %that can lead
{leading} to non-classical correlations between these two degrees of freedom {(intra-photon entanglement~\cite{azzini_single-particle_2020})}. {We perform time resolved polarization tomography that allow us to measure the  density matrix of the intra-entangled state and characterize the photonic qubit source in terms of efficiency, single-photon purity and indistinguishability}.

\begin{figure}
    \includegraphics[width=1\linewidth]{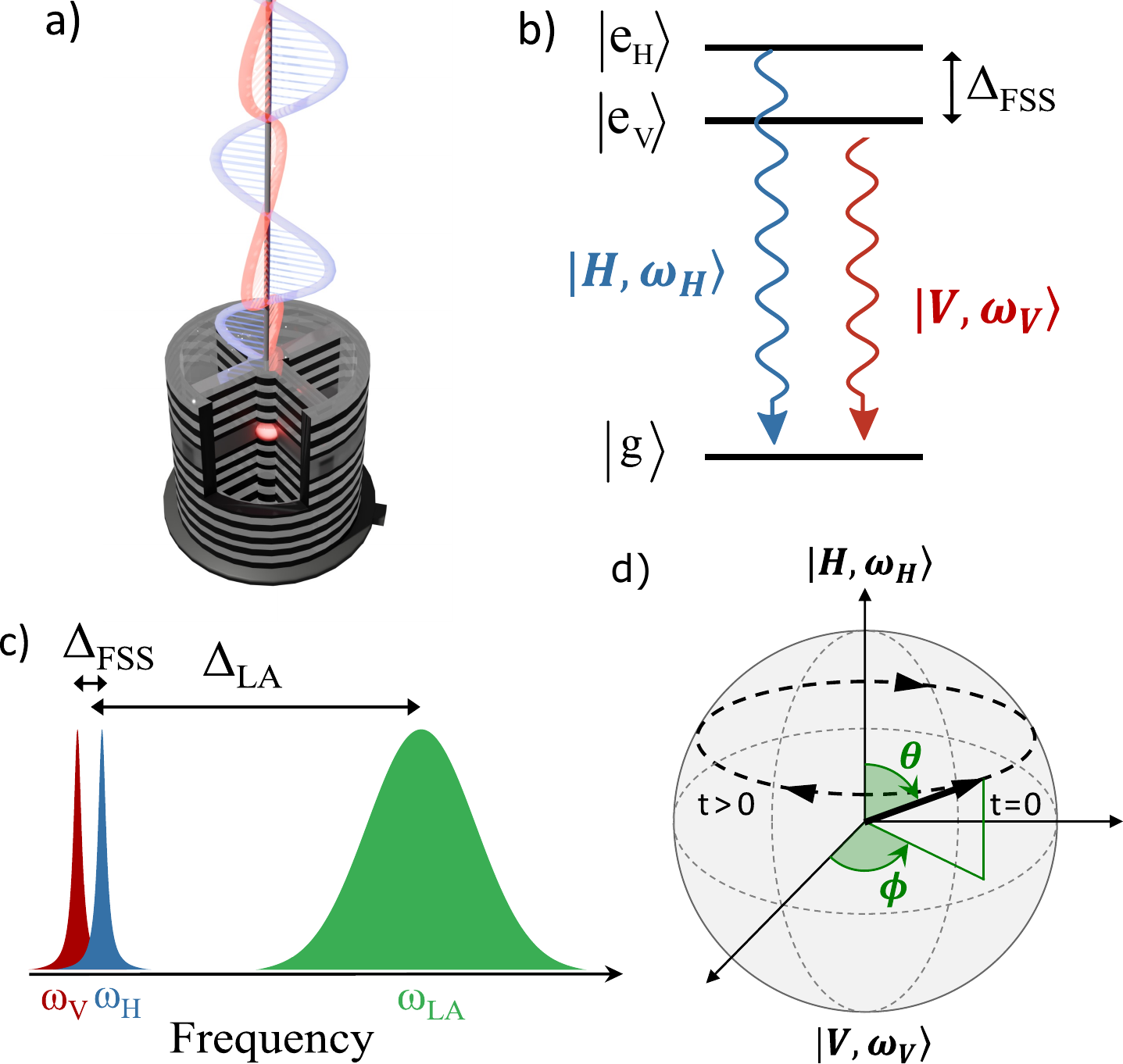}
    \caption{ 	{Principle of the experiment. a) Artistic visual of the device, a semiconductor quantum dot (red) embedded in a micropillar cavity, generating a photonic qubit hyper-encoded in the frequency and polarization degrees of freedom. b) Energy levels of the neutral QD, consisting in one ground state and two excited states separated in energy by the fine structure splitting $\Delta_{FSS}$ and emitting orthogonal, linearly polarized photons. c) Sketch of the excitation technique. A near-resonant laser pulse {(green)} with detuning $\Delta_{LA} \gg \Delta_{FSS}$ can excite the two dipole transitions via a phonon-assisted process, either individually or in a superposition. d) Bloch sphere representation of the generated photonic state. The parameters $\theta$ and $\phi$ {(vector at $t=0$)} are controlled by the pump laser polarization. The energy difference between the two states $\Delta_{FSS}$ induces a coherent evolution of the photonic state for $t>0$.
    } \label{Fig1}}
\end{figure}

The device under study is illustrated in figure \ref{Fig1}a. It consists of an InGaAs QD deterministically embedded in a micropillar structure surrounded by Distributed Bragg Reflectors~\cite{nowak_deterministic_2014}, and electrically contacted~\cite{ollivier_reproducibility_2020}. The device is placed in a closed-cycle cryostat and cooled down to 8K. Single photons are generated and collected using a confocal microscopy setup~\cite{thomas_bright_2021}. Figure \ref{Fig1}b shows the energy levels of the neutral QD, consisting in a {three-level} structure with one ground state $\ket{g}$ and two excited states $\ket{e_H}$ and $\ket{e_V}$ separated in energy by the fine structure splitting ($\Delta_{FSS}$). This splitting of typically few to tens of $\mu$eV in InGaAs QDs arises from the natural asymmetry in self-assembled quantum dots~\cite{bayer_fine_2002} due to the strain-induced growth process. Another consequence of this anisotropy is that the two optical transitions occur with linear, orthogonal polarizations denoted $\ket{H}$ and $\ket{V}$. In this work, we make use of these two dipole transitions as the eigenmodes of a photonic qubit.

We generate single photons from the QD using the Longitudinal Acoustic (LA)-phonon-assisted excitation scheme~\cite{barth_fast_2016,thomas_bright_2021} depicted in figure \ref{Fig1}c. In this scheme, a 15ps long laser pulse, blue-detuned from the QD central wavelength %at 924.9nm 
by $\Delta_{LA}$=0.65nm, can excite both dipoles with similar efficiencies since {$\Delta_{LA}$ $\gg$ $\Delta_{FSS}$}. The strong excitation pulse first dresses the ground and excited states, enabling the fast (few ps) relaxation of a LA phonon between the dressed states. Adiabatic undressing of the energy levels at the end of the laser pulse then leaves the QD in its excited state with high probability, which then recombines through spontaneous emission of a photon. Despite being an incoherent process, this scheme has been demonstrated to yield an efficient population inversion along with high levels of single photon purity and indistinguishability~\cite{thomas_bright_2021}. It also faithfully preserves the polarization of the selection rules and its broadband nature allows  exciting both dipoles~\cite{reindl_highly_2019}, a crucial feature for this work. The detuning $\Delta_{LA}$ between the pump laser pulse and the QD then allows to spectrally filter the pump laser pulse, allowing to excite and collect photons with arbitrary polarization. In particular, the polarization of the pump laser can be set either to address selectively one of the two transitions, or to generate a superposition of the two dipoles.

The general case is depicted in figure \ref{Fig1}d. The excitation laser pulse has polarization $\ket{\theta,\phi} = \cos (\theta) \ket{H} + \sin (\theta) \mathrm{e} ^ { \im \phi } \ket{V}$, where $\theta$ is the angle between the laser polarization and the $H$ dipole of the QD, and $\phi$ accounts for the ellipticity of the pulse ($\phi = 0$ corresponding to linear polarization). {This pulse prepares the state $\ket{\psi_{QD}}$ of the QD in a superposition of the two excited states $\ket{e_H}$ and $\ket{e_V}$ corresponding respectively to the horizontally and vertically polarized dipoles,} at $t=0$ which evolves {until decay according to the quantum trajectory} \begin{equation}
    \ket{\psi_{QD}(t)} = \begin{aligned}[t] 
    & \cos (\theta) \mathrm{e} ^ { - \im \omega_H t } \mathrm{e} ^ {- \frac{t}{2\tau}} \ket{e_H} \\
    & + \sin (\theta) \mathrm{e} ^ { - \im (\omega_V t - \phi) } \mathrm{e} ^ {- \frac{t}{2\tau}} \ket{e_V}
\end{aligned} 
\label{eqn_state_evol}
\end{equation}
{where $\omega_H$ ($\omega_V$) is the central frequency of the horizontally (vertically) polarized dipole and $\tau$ is the lifetime of the excited state.} This temporal evolution of the excited state of the QD is then imprinted on the state of the emitted photon $\ket{\psi}$ {as a consequence of the proportionality relation between the QD dipole operator and the field operator of the electromagnetic mode. In the absence of dephasing, the photonic state can thus be written as
\begin{equation}
    \ket{\psi} = 
    \cos (\theta)\ket{H,\omega_H} + \sin (\theta)e^{\im\phi}\ket{V,\omega_V}
\label{eqn_state}
\end{equation}
where the pulse mode $\ket{\omega_j}$ is
\begin{equation}
    \ket{\omega_j}=\frac{1}{\sqrt{\tau}}\int_0^\infty \mathrm{e}^{-\im\omega_j t}\mathrm{e}^{-\frac{t}{2\tau}}\hat{a}_j^\dagger(t)\ket{0}dt
\label{eqn_freqstates}
\end{equation}
in the time basis, for polarization $j\in\{H,V\}$.
}

\begin{figure*}
    \includegraphics[width=1\linewidth]{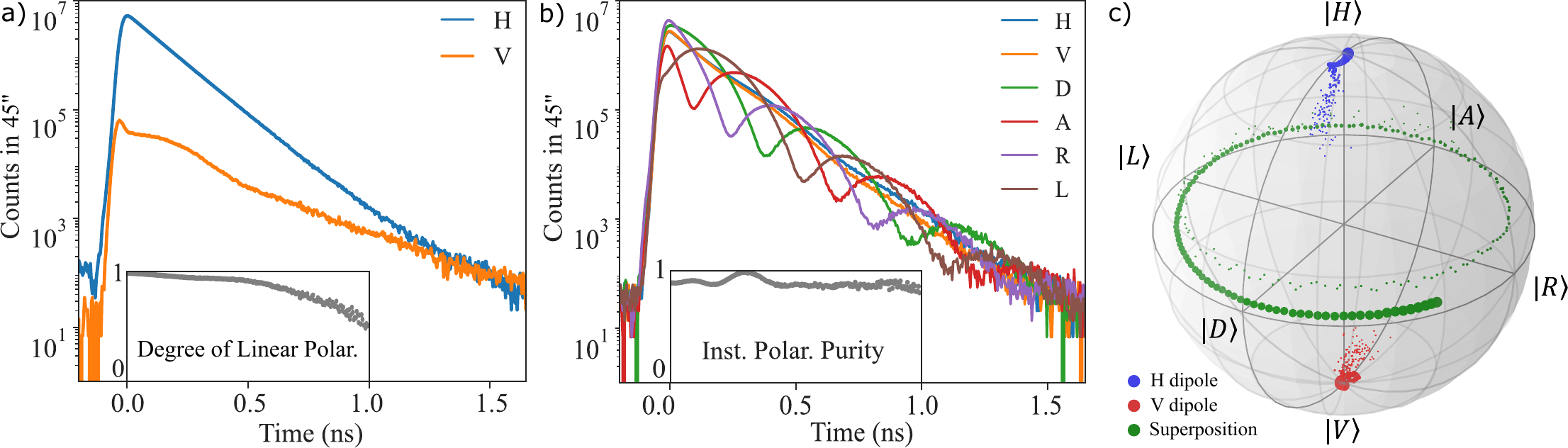}
    \caption{ 	{Time-resolved polarization tomography of the generated quantum states. a) Time traces for a pure state (H dipole) measured in the H/V basis, and b) for a frequency-polarization hyper-encoded photonic qubit, measured in all 3 bases (full tomography). The insets show respectively the degree of linear polarization and instantaneous purity over time between 0 and 1ns (linear scale). c) Trajectory of the photonic qubits in the Bloch sphere. Points sizes are weighted by their intensity. 
    } \label{Fig_lifetimes}}
\end{figure*}

{We remark that, provided $\Delta\omega=\omega_H-\omega_V=\Delta_{FSS}/\hbar$ is nonzero, the state in equation (\ref{eqn_state}) is non-separable}: the frequency and polarization degrees of freedom of a single emitted photon are entangled, a feature referred to as \textit{intra-particle} entanglement~\cite{suprano_orbital_2022}, or \textit{single-particle} entanglement~\cite{azzini_single-particle_2020}. Such a feature can be exploited in quantum information protocols for {instance in robust} quantum communication~\cite{cozzolino_high-dimensional_2019}, {efficient quantum logic}~\cite{lanyon_simplifying_2009}~\cite{imany_high-dimensional_2019}, or random number generation~\cite{leone_certified_2022}. Alternatively, it can be removed in a deterministic manner~\cite{fiorentino_deterministic_2004}. In the following, we characterize the {polarization and frequency hyper-encoding} by performing {polarization-resolved} measurements in the time domain. For this, we use Superconducting Nanowire Single Photon Detectors (SNSPDs) with temporal resolution below $30$ps {$\ll{2\pi}/\Delta\omega$}, allowing us to resolve the temporal evolution described in equation \ref{eqn_state_evol}.

We first generate single photons in a pure state for each dipole% (similar to ref.\cite{thomas_bright_2021})
{, $\ket{H, \omega_H}$ or $\ket{V, \omega_V}$}. The emitted photons are sent through a polarization tomography stage and detected by SNSPDs. We set the polarization of the exciting laser parallel to the H dipole and record the time traces of the emission in polarization parallel (H) and perpendicular (V) to the dipole. The obtained time traces are shown in figure \ref{Fig_lifetimes}a. We observe a mono-exponential decay, giving a {radiative lifetime of $\tau = 133$ps}, highly polarized along the dipole polarization. The inset displays the degree of linear polarization ($D_{LP}$) of the emission over time, between 0 and 1ns, (or equivalently, the Stokes parameter in the HV basis, $S_{HV}$) defined as:

\begin{equation}
    D_{LP} = S_{HV} = \frac{I_H - I_V}{I_H + I_V} \label{eqn_DLP}
    \end{equation}

Integrating over the full wavepacket, we measure $D_{LP} = 97.1\%$, while the instantaneous $D_{LP}$ slightly decreases over time indicating the presence of dissipative processes in the excited state. We perform the same measurement with the pump laser aligned along the V dipole ({data in supplementary}), measuring a time average $D_{LP} = 96.8\%$ for this dipole.

We now set the polarization of the pump laser to excite the two dipoles with equal probability, and record the time traces of the emission, this time %in all 6 polarization bases 
{along all 6 polarization axes} (full polarization tomography). The results are displayed in figure \ref{Fig_lifetimes}b. We observe monoexponential decays in the H and V bases with equal intensities, whereas coherent oscillations appear the D/A/R/L {polarizations}. Such time traces can be understood by projecting the photonic wavepacket of equation {(\ref{eqn_state})} along the corresponding polarization measurement bases. In particular, performing a measurement along the D basis gives the measured intensity:

{
\begin{IEEEeqnarray}{rCl}
    {I_D(t)} 
    % &=&  |\braket{\sw{D(t)|\psi}}|^2 \nonumber \\
    &=&  \frac{1}{2}\braket{\psi|(\hat{a}_H(t)+\hat{a}_V(t))^\dagger(\hat{a}_H(t)+\hat{a}_V(t))|\psi} \nonumber \\
    &=&  \frac{1}{2\tau}\mathrm{e} ^ {-\frac{t}{\tau}} \left(1+\cos \left( \Delta\omega t - \phi \right)\sin(2\theta)\right)  
    \label{eqn_ID}
\end{IEEEeqnarray}}

Similar expressions can be derived for the A/R/L bases, whereas the time trace following a projective measurement in the eigenstates bases H and V result only in a monoexponential decay. From equation \ref{eqn_ID}, we deduce that the period of the oscillations provides a direct measurement of the fine structure splitting, and obtain via a fit of the time trace $\Delta_{FSS}=\hbar\Delta\omega =7.35~\mu$eV. This  value is consistent with values obtained for annealed QDs studied here where strain and anisotropy is reduced.

Using the time traces from figure \ref{Fig_lifetimes}, we reconstruct the Stokes vector of the state over time, defined as $S = (S_{HV}, S_{DA}, S_{RL})$ with $S_{DA}$ and $S_{RL}$ defined similarly as in equation \ref{eqn_DLP}. We plot the state evolution over time in the {Bloch} sphere in figure \ref{Fig_lifetimes}c (green points), where we observe a coherent evolution around the H/V axis, as described in figure \ref{Fig1}d. For comparison, in blue {data from figure \ref{Fig_lifetimes}a} and red {data in supplementary} are the eigenstates generated when exciting along the dipoles H and V. 

The inset in figure \ref{Fig_lifetimes}b shows the instantaneous polarization purity over time, defined as the norm of the Stokes vector. {A high purity is observed with    hardly any decrease over the lifetime, with an integrated purity of $90.0\%$ over the total wavepacket.}

\begin{figure}
    \includegraphics[width=1\linewidth]{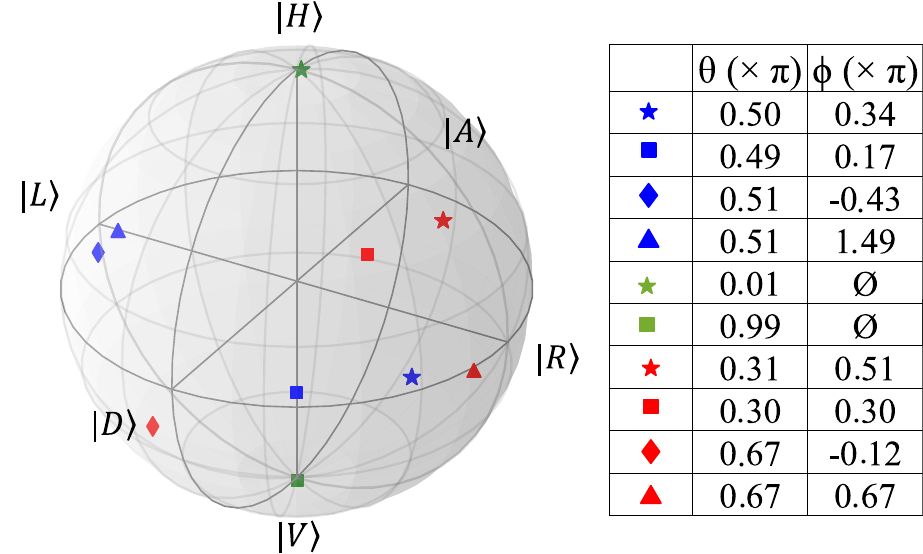}
    \caption{ 	{Generation of any arbitrary superposition between the two dipoles. The laser polarization is varied to generate 10 states with different initial amplitudes and phases. The fits of the time-resolved polarization tomography for each state allow to extract the values of $\theta$ and $\phi$, which we display in the table.} \label{Fig_superpositions}}
\end{figure}

To demonstrate that we can generate such state with arbitrary initial amplitude and phase, we now vary the pump laser polarization (angles {$\theta_{laser}$ and $\phi_{laser}$}) and perform similar time-resolved polarization tomography measurements for 10 different sets of excitation polarization angles in total. From the time traces, we retrieve the values of {the angles $\theta$ and $\phi$ for the generated state} at $t=0$, which we display in the Bloch sphere in figure \ref{Fig_superpositions} along with the numerical values. The points span across the entire Bloch sphere, including equal and unequal superpositions with arbitrary initial phase. {We note that in the general case, the photonic state is described by a vector in a composite Hilbert space with 2 dimensions of frequency and 2 of polarization, whose basis is constituted by the 4 possible single-photon states $\ket{H, \omega_H}$, $\ket{V, \omega_V}$, $\ket{H, \omega_V}$ and $\ket{V, \omega_H}$ . While our scheme allows to generate arbitrary superpositions of the first two basis states, the other two are accessible by simple unitary operations on the polarization of the generated photons, i.e. with a set of waveplates.}

\begin{figure}
    \includegraphics[width=1\linewidth]{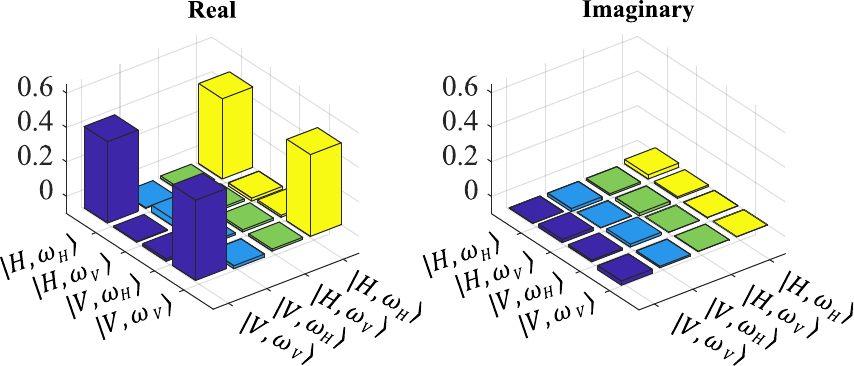}
    \caption{Intra-entanglement between the polarization and frequency degrees of freedom of a single photon : Reconstructed density matrix of a hyper-encoded state in frequency and polarization. The fidelity to the target state is {estimated at} $F = 94\pm8\%$.
    \label{Fig_densitymatrix}}
\end{figure}

We now turn to the analysis of the single-particle entanglement arising from the non-separability between polarization and frequency in equation {(\ref{eqn_state})}. Using the polarization-resolved time traces (figure \ref{Fig_lifetimes}b), we can retrieve the coefficients of the density matrix of the generated state {in the composite Hilbert space}. We derive the expressions of the expected intensities over time after projecting the wavepacket described in equation {(\ref{eqn_state})} along the 6 measurement {axes} (detailed in supplemental information), and obtain a set of equations which we fit to the time traces, using a maximum likelihood method \cite{james_measurement_2001}. This allows us to estimate the coefficients of the density matrix, which we show in figure \ref{Fig_densitymatrix}. The obtained density matrix is close to the expected one for the {target} state $\ket{\phi^+} = \frac{1}{\sqrt 2} \ket{H, \omega_H} + \ket{V, \omega_V}$, with {estimated fidelity $F = 94\pm8\%$}. {Our approach could also generate the orthogonal Bell state $\ket{\phi^-} = \frac{1}{\sqrt 2} \ket{H, \omega_H} - \ket{V, \omega_V}$ by exciting with the orthogonal polarization, while the other two Bell states $\ket{\psi^\pm} = \frac{1}{\sqrt 2} \ket{H, \omega_V} \pm \ket{V, \omega_H}$ can be obtained by combining the scheme with polarization unitary operations.}

Finally, we assess the performance of the device as single photon source, namely in terms of efficiency, single-photon purity and indistinguishability. In figure \ref{Fig_lifetimes}a, the total detected rate for the H dipole is 4.7 MHz, for a repetition rate of the laser $R_\mathrm{L} = 82$~MHz. Considering optical losses in our experimental setup (see supplementary), this corresponds to a first lens brightness value of $\mathcal{B}_\mathrm{FL} = 28 \pm 2\%$. A slightly lower detected count rate of 3.4 MHz is found for the V dipole, (data in supplementary) corresponding to a first lens brightness of $21 \pm 2\%$. Note that several QD based devices have already demonstrated brightness values above $50\%$~\cite{tomm_bright_2021}, including a first-lens brightness of $55\%$ in the LA phonon-assisted scheme~\cite{maring_versatile_2024}, setting an estimation of the achievable rates.

\begin{figure}
    \includegraphics[width=0.9\linewidth]{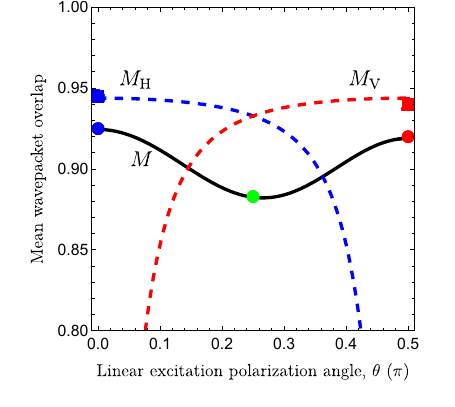}
    \caption{ 	{Mean wavepacket overlap M obtained from the HOM interference visibility, measured for the H (blue, $g^{(2)}(0) = 1.47\%, M = 91.6 \pm 0.1\%$) and V (red, $g^{(2)}(0) = 1.47\%, M = 92.2 \pm 0.1\%$) dipoles, as well as for an equal superposition of the two dipoles (green, $g^{(2)}(0) = 1.7\%, M = 88.3 \pm 0.1\%$). The circles are measured without polarization projection, while the squares are the values $M_H$ and $M_V$ taken after the photons are filtered with a polarizer. The black line shows the expected value from our model (see text), {and the dashed lines correspond to the simulated values of $M_H$ and $M_V$}}. 
    } \label{Fig_g2_hom}
\end{figure}

The second-order intensity correlation gives $g^{(2)}(0)$ as low as $g^{(2)}(0)=1.70 \pm 0.02\%$ and $g^{(2)}(0)=1.47 \pm 0.01\%$ respectively for each dipole, and $g^{(2)}(0)=1.3 \pm 0.1\%$ for a balanced superposition of the two dipoles (see supplementary data), demonstrating the single-photon nature of the source. The indistinguishability is measured via the Hong-Ou-Mandel interference of the full wavepacket, without filtering the photons with a polarizer. We obtain respectively $V_{HOM} = 88.9 \pm 0.3\%$ and $88.8 \pm 0.3\%$ for the pure dipoles H and V, corresponding to indistinguishabilities of $M = 92.2 \pm 0.1\%$ and $91.6 \pm 0.1\%$ \cite{ollivier_hong-ou-mandel_2021} as shown in figure \ref{Fig_g2_hom}, which further increase to $M_H = 94.5 \pm 0.1\%$ and $M_V = 93.8\pm 0.1\%$ if the photons are filtered with a polarizer. For a balanced quantum superposition of the two dipoles, we measure the full wavepacket overlap, comprised of the two orthogonaly polarized modes. We obtain $V_{HOM} = 85.9 \pm 0.1\%$, corresponding to $M = 88.3 \pm 0.1\%$. This value is only slightly lower than the one obtained for {the eigenstates}, in line with the integrated polarization purity (inset of figure \ref{Fig_lifetimes}) and further evidences the coherence of the state generated.

{The dashed lines and the black line in figure \ref{Fig_g2_hom} display} the expected interference visibilities predicted by our model, taking into account {a dissipative mixing of the $\ket{e_H}$ and $\ket{e_V}$ excited states, a slight imbalance between the two {measured} dipole magnitudes, and an initial state preparation impurity (see supplemental). {The dashed lines show that the polarization-filtered visibilities drop substantially as the excitation polarization becomes orthogonal to the filtered polarization. This is caused by the filter progressively allowing proportionally more photons through that arise following a polarization flip in the QD excited state, and hence the light becomes more incoherent. The fit of the black line to the unfiltered visibility measurements allows us to estimate the timescale of this dissipative mixing to be $12\pm1$ns.}  The difference between the unfiltered values at $\theta=0$ and $\theta=\pi/2$ is explained by a V-polarized dipole that is about {12\% weaker}, which qualitatively agrees with the observed difference in the count rates. We also find that the reduced mean wavepacket overlap given by the superposition state implies an initial state preparation purity of $94.9\pm0.2\%$, which corresponds to the observed non-unity instantaneous polarization purity at $t_0$ (see inset of figure \ref{Fig_lifetimes}b).}

An important parameter in our approach is the QD fine structure splitting, which governs {the overlap between states $\ket{\omega_H}$ and $\ket{\omega_V}$ by $|\braket{\omega_H|\omega_V}|^2=1/(1+\Delta\omega^2 \tau^2)$ following equation (\ref{eqn_freqstates}).} The relatively small {splitting} in our devices based on annealed QDs results in a non negligible  overlap between the two dipoles: considering the radiative lifetime of 133ps and the energy splitting of $\Delta_{FSS}=7.35 \mu$eV, we calculate an overlap of {$31.2\%$. We have experimentally verified} this value via the HOM interference between the two dipoles~\cite{legero_quantum_2004} (see supplementary), for which we obtained a value of $M_\omega = 31.8 \pm 0.1\%$ in good agreement. {The overlap also dictates the amount of intra-particle entanglement. By applying a Gram-Schmidt orthogonalization to equation (\ref{eqn_state}) (see supplementary), we find that the polarization-frequency entanglement concurrence is bounded from above by $C\leq \sqrt{1-|\braket{\omega_H|\omega_V}|^2}\simeq 83\%$. Using the measured overlap from HOM interference $|\braket{\omega_H|\omega_V}|^2\simeq M_\omega$, and by applying the same orthogonalization process to the reconstructed density matrix, provides an entanglement concurrence of $C=77\%$.}

An FSS of 200$~\mu$eV, that is %commonly observed for as grown
{reasonably achievable for} InAs QDs would result in an overlap {of 0.06\%}, allowing to properly resolve the two frequency modes {and providing an entanglement concurrence as high as $99.97\%$}. Moreover, we note that the FSS of a quantum dot can be made tunable by engineering the device. This has been an extensive topic of research in the context of entangled photon pair generation with neutral QDs~\cite{schimpf_quantum_2021}~\cite{jons_bright_2017}, and was demonstrated using strain~\cite{huber_strain-tunable_2018}, magnetic field~\cite{stevenson_semiconductor_2006}, or in-plane electric fields~\cite{ollivier_three-dimensional_2022}.

In conclusion, we have reported {on an efficient approach to} generate hyper-encoded single photons in a superposition of their frequency and polarization degrees of freedom {in a deterministic manner}. Our device, a semiconductor quantum dot in a microcavity, brings near-deterministic single photon emission at high rates to the field of frequency encoding. The emitted single photons show high purity and indistinguishability, enabling them {to perform quantum interferences for quantum information processing applications}. 

{Going further, the present scheme could be transferred to other QD materials, such as droplet GaAs QDs for interfacing with quantum memories~\cite{zaporski_ideal_2023}, or InP QDs for telecommunication wavelength operation~\cite{yu_telecom-band_2023}. We emphasize that} the inhomogeneous distribution in frequency between various QDs could be leveraged to reach a large number of frequency modes, by combining the present scheme with several QD devices and frequency gates~\cite{lukens_frequency-encoded_2017}. Finally, the scheme could be extended to generate {spin-entangled frequency qudits~\cite{adambukulam_hyperfine_2024}} using the four non-degenerate optical transitions of a charged quantum dot in a high magnetic field, or even to larger dimensions using magnetic quantum dots~\cite{baudin_optical_2011}.

\noindent \textbf{Acknowledgements.}  The authors thank Andrew White for valuable discussions. This work was partially supported the European Union's Horizon 2020 FET OPEN projects  QLUSTER (Grant ID 862035) and PHOQUSING (grant ID 899544), Horizon-CL4 program under the grant agreement 101135288 for EPIQUE project, the French RENATECH network, by the European Commission as part of the EIC accelerator program under the grant agreement 190188855 for SEPOQC project, the Plan France 2030 through the projects ANR-22-PETQ-0011 and ANR-22-PETQ-0013.
N.C. acknowledges support from the Paris Ile-de-France R\'egion in the framework of DIM SIRTEQ. S.C.W. acknowledges support from the Foundational Questions Institute Fund (Grant No. FQXi-IAF19-01). {S.E.E. acknowledges supported from the  from the NSF (Grant No. 1741656)}
\vspace{0.5cm}
\bibliography{Biblio-DF-NC}

\end{document}

\title{Supplementary material}

\maketitle

\section{Supplementary data}

\begin{figure}[h]
    \centering
    \includegraphics[scale=0.9]{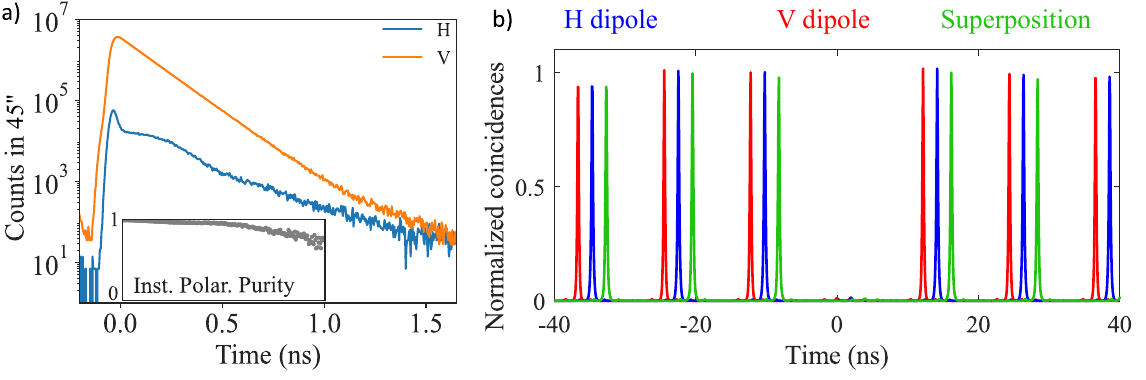}
    \caption{Supplementary data. a) Polarization-resolved lifetime measurement of the V dipole measured in the H/V basis. The inset shows the degree of linear polarization over time between 0 and 1 ns. b) $g^{(2)}$ measurements of the polarization-unfiltered photonic wavepacket (circles from figure 5 of the main text). Graphs have been offset horizontally for clarity.}
    \label{figS_extradata}
\end{figure}

 Figure \ref{figS_extradata}a shows the polarization-resolved lifetime measurement of the V dipole measured in the H/V bases. The result is similar to the case of the H dipole (figure 2a), only with a slightly lower count rate, with a detected rate $R_\mathrm{det} = 3.4$MHz compared to $4.7$MHz for the H dipole. The inset shows the degree of linear polarization ($D_{LP}$) over time between 0 and 1 ns. The integrated $D_{LP}$ is $96.8\%$. Figure \ref{figS_extradata}b shows the histograms of the second order intensity correlation performed to measure the single photon purity. The $g^{(2)}$ is $1.70 \pm 0.02\%$ and $1.47 \pm 0.01\%$ respectively for each dipole, and $1.3 \pm 0.1\%$ for a balanced superposition of the two dipoles

\section{Experimental setup and loss budget}

The experimental setup for single photon excitation and collection is shown in fig. S2. To generate the excitation laser pulses, we start from a 3ps pulsed Ti:Sapphire laser with a repetition rate of 82MHz and a central wavelength blue-detuned by 0.65nm from the quantum dot transition, which emits at a wavelength of 924.9nm$\pm \frac{\Delta_{FSS}}{2}$. The laser spectrum is shaped using a 4-f zero-dispersion line with a slit in the Fourier plane to obtain pulses with the desired length of 15 ps. The detuning between the QD central wavelength and the laser pulses ensures that filtering of the laser can be done by means of 3 high transmission, narrowband (0.8nm bandwidth) spectral filters. These filters also serve as dichroïc mirrors, reflecting the laser pulse from the excitation arm towards the quantum dot path, and transmitting only the emitted photons towards the collection path. The polarization of the excitation laser is controlled by motorized waveplates (half and quarter) in the excitation arm. Additional waveplates are located on top of the cryostat to compensate for birefringence of the various optical components in the setup. The laser pulse is focused on the micropillar using a high NA (0.7) aspheric lens inside the cryostat. The collected single photons can be sent to either a polarization tomography stage, a Hanbury-Brown-Twiss (HBT) setup or a Hong-Ou-Mandel (HOM) interferometer, before being detected by Superconducting Nanowire Single Photon Detectors (SNSPDs) with $<30$ ps timing jitter and an efficiency of 80\% recorded at the detection rate of 5MHz. The intensity time traces of the emission, and correlations measurements are recorded using a HydraHarp correlator, for a total temporal jitter of our detection of 30ps FWHM.

\begin{figure}[h]
    \centering
    \includegraphics[scale=0.9]{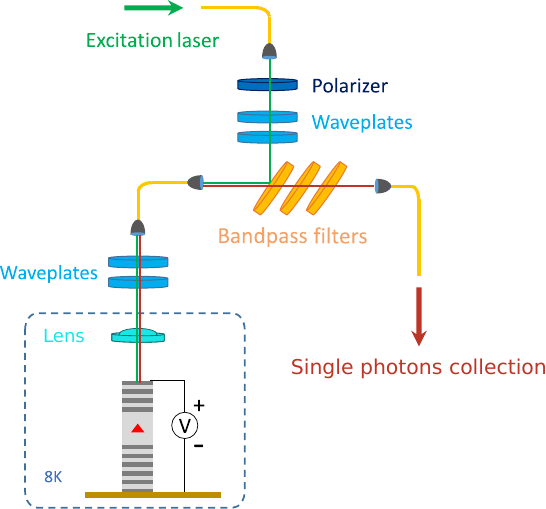}
    \caption{Experimental setup for single photon generation and collection using LA-phonon assisted excitation. The device is placed in a cryostat at 8K. The polarization of the laser pulse is controlled in the excitation arm by a set of polarizer and waveplates to generate the desired superposition of the two dipole transitions.}
    \label{fig_setup}
\end{figure}

The transmission of each component in the single photon collection path is measured using a continuous wave laser centered at the QD wavelength. The table S1 summarizes the measured values and yields a transmission efficiency from the first lens until (including) the collection single mode fiber of $T_{setup} = 32 \pm 2\%$. Similarly, the transmission of each component in the tomography stage is given in table S2 with a total transmission $T_{tom} = 78 \pm 2\%$. The efficiency of the SNSPDs $\eta_{det}$ is measured with attenuated laser pulses from a pulsed laser with a repetition rate of 82MHz, at $\eta_{det} = 80 \pm 3\%$. Given the single photons detection rates of $R_\mathrm{det} = 4.7$~MHz and 3.4MHz respectively for the H and V dipoles for a repetition rate of the laser $R_\mathrm{L} = 82$~MHz we obtain for the first lens brightness $\mathcal{B}_\mathrm{FL} = R_\mathrm{det}/(R_\mathrm{L}\times T_{setup} \times T_{tom} \times \eta_{det}) = 28 \pm 2 \%$ and $21 \pm 2 \%$ respectively for the H and V dipoles, and the corresponding fibered brightness $\mathcal{B}_\mathrm{Fib} = \mathcal{B}_\mathrm{FL} \times T_{setup} = 10 \pm 1\%$ and $7 \pm 1\%$ respectively.

\begin{table}[h]
\begin{center}
\begin{tabular}{|c|c|}
\hline
     Element & Transmission  \\
     \hline
     Lens + Cryostat Window & $0.90 \pm 0.01$  \\
     QWP + HWP & $0.98 \pm 0.01$ \\
     Free-space to fiber coupling  & $0.65 \pm 0.03$ \\
     Fiber transmission & $0.92 \pm 0.01$\\
     BP filters x3 & $0.82 \pm 0.01 $\\
     Free-space to fiber coupling & $0.83 \pm 0.01$ \\
     %Polarization tomography stage & $0.78 \pm 0.01$ \\
     Fiber transmission & $0.90 \pm 0.01$ \\
     \hline
     Total & $0.32 \pm 0.02$ \\
     \hline
\end{tabular}
\end{center}
\caption{Transmission of the optical elements in the setup. QWP: Quarter Waveplate. HWP: Half Waveplate. \label{TableS1} }
\end{table}

\begin{table}[h]
\begin{center}
\begin{tabular}{|c|c|}
\hline
     Element & Transmission  \\
     \hline
     QWP x2 + HWP x2 & $0.96 \pm 0.01$ \\
     Wollaston prism  & $0.97 \pm 0.01$ \\
     Free-space to fiber coupling & $0.88 \pm 0.01$\\
     Fiber transmission & $0.95 \pm 0.01 $\\
     \hline
     Total & $0.78 \pm 0.01$ \\
     \hline
\end{tabular}
\end{center}
\caption{Transmission of the optical elements in the polarization tomography stage. QWP: Quarter Waveplate. HWP: Half Waveplate. \label{TableS2} }
\end{table}

\section{Measurement of the spectral overlap between the two dipoles}

\begin{figure}[h]
    \centering
    \includegraphics[scale=0.45]{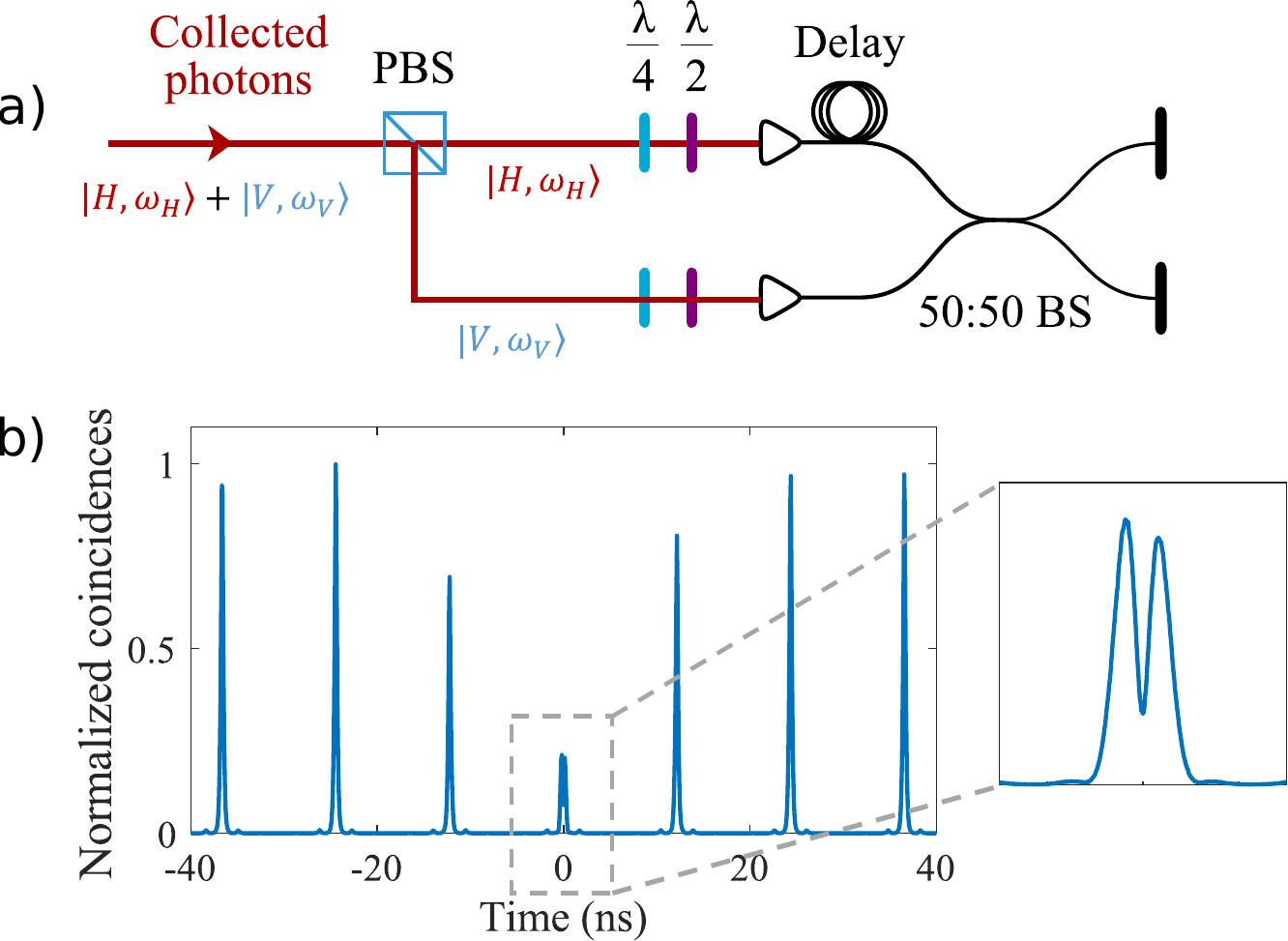}
    \caption{Measurement of the spectral overlap between the two dipoles via a Hong-Ou-Mandel interference measurement. a. Description of the interferometer: a hyperencoded single photon is sent to a PBS so that each frequency mode is sent to a different arm of the interferometer. b. Measured HOM visibility. The inset displays the central peak.}
    \label{fig_measure_spectral_overlap}
\end{figure}

In this section, we experimentally verify the calculated value of the overlap between the two dipoles by using the HOM interference between two photons generated by each dipole. As illustrated in figure \ref{fig_measure_spectral_overlap}a, a hyperencoded single photon is generated and the two frequency components are separated via their polarization and one arm is delayed in order to measure interference between successive photons. We then interfere the two dipoles in a HOM setup and measure the visibility of the interference. The histogram for this experiment is displayed in figure \ref{fig_measure_spectral_overlap}b, from which we obtain $M_\omega = 31.8 \pm 0.1\%$, in good agreement with the predicted value of $31.2\%$. The inset displays the central peak of the histogram, showing the effect of the frequency mismatch between the two photons~\cite{legero_quantum_2004}, consistent with the value of $\Delta_{FSS} = 7.35 \mu eV$.

\section{State tomography of a single photon in the polarization and frequency basis}

In this section we %detail the method used to 
reconstruct the density matrix of the frequency-polarization hyperencoded single photon state. We consider a composite Hilbert space with 2 dimensions of frequency and 2 of polarization, whose basis is constituted by the 4 possible single-photon states $\ket{H, \omega_H}, \ket{H, \omega_V}, \ket{V, \omega_H}, \ket{V, \omega_V}$. We treat the problem in the Heisenberg representation, where the state vector is fixed and the operators are time dependent. The time evolution of a pure quantum state $\ket{\Psi}$ in this picture is written as: 
{
\begin{equation}
    \ket{\Psi} = a \ket{H, \omega_H} + b \ket{H, \omega_V} + c \ket{V, \omega_H} + d \ket{V, \omega_V} \\
    \end{equation}}
with $a, b, c, d \in \mathbb{C}$. In general, the density matrix of a mixed quantum state $\rho$ in the considered Hilbert space has the form
\begin{equation}
\rho = \kbordermatrix{
& \bra{H, \omega_H} & \bra{H, \omega_V} & \bra{V, \omega_H} & \bra{V, \omega_V} \\
\ket{H, \omega_H} & \rho_{11} & \rho_{12} & \rho_{13} & \rho_{14} \\
\ket{H, \omega_V} & \rho_{21} & \rho_{22} & \rho_{23} & \rho_{24} \\
\ket{V, \omega_H} & \rho_{31} & \rho_{32} & \rho_{33} & \rho_{34} \\
\ket{V, \omega_V} & \rho_{41} & \rho_{42} & \rho_{43} & \rho_{44} \\ }
\end{equation}
with complex coefficients $\rho_{i,j}$. We now define the projection operators along the 6 polarization bases:
\begin{equation}
A_H = \ket{H}\bra{H} \nonumber
\end{equation}
and similar for D, A, R and L
\begin{equation}
\ket{H} = \ket{\mathrm{e}^{i\Delta\omega t/2}, \mathrm{e}^{-i\Delta\omega t/2}, 0, 0}, \ket{V} = \ket{0, 0, \mathrm{e}^{i\Delta\omega t/2}, \mathrm{e}^{-i\Delta\omega t/2}} \nonumber
\end{equation}
and
\begin{equation}
\ket{D} = \frac{1}{\sqrt{2}}(\ket{H} + \ket{V}), \ket{A} = \frac{1}{\sqrt{2}}(\ket{H} - \ket{V}),
\ket{R} = \frac{1}{\sqrt{2}}(\ket{H} - i \ket{V}), \ket{L} = \frac{1}{\sqrt{2}}(\ket{H} + i \ket{V}) \nonumber
\end{equation}

This leads to the following projection operators:
\begin{center}
\begin{equation}
    \begin{aligned}[t]
    & A_H =
    \begin{pmatrix}
        \mathrm{e}^{i\Delta\omega t} & 0 & 0 & 0 \\
        0 & \mathrm{e}^{-i\Delta\omega t} & 0 & 0 \\
        0 & 0 & 0 & 0 \\
        0 & 0 & 0 & 0 \\
        \end{pmatrix},
    A_V =
    \begin{pmatrix}
        0 & 0 & 0 & 0 \\
        0 & 0 & 0 & 0 \\
        0 & 0 & \mathrm{e}^{i\Delta\omega t} & 0 \\
        0 & 0 & 0 & \mathrm{e}^{-i\Delta\omega t} \\
        \end{pmatrix}, \\
    & A_D =
    \begin{pmatrix}
        1/2 & 1/2 \mathrm{e}^{i\Delta\omega t} & 1/2 & 1/2 \mathrm{e}^{i\Delta\omega t} \\
        1/2 \mathrm{e}^{-i\Delta\omega t} & 1/2 & 1/2 \mathrm{e}^{-i\Delta\omega t} & 1/2\\
        1/2 & 1/2 \mathrm{e}^{i\Delta\omega t} & 1/2 & 1/2 \mathrm{e}^{i\Delta\omega t} \\
        1/2 \mathrm{e}^{-i\Delta\omega t} & 1/2 & 1/2 \mathrm{e}^{-i\Delta\omega t} & 1/2\\
        \end{pmatrix}, 
    A_A =
    \begin{pmatrix}
        1/2 & 1/2 \mathrm{e}^{i\Delta\omega t} & -1/2 & -1/2 \mathrm{e}^{i\Delta\omega t} \\
        1/2 \mathrm{e}^{-i\Delta\omega t} & 1/2 & -1/2 \mathrm{e}^{-i\Delta\omega t} & -1/2\\
        -1/2 & -1/2 \mathrm{e}^{i\Delta\omega t} & 1/2 & 1/2 \mathrm{e}^{i\Delta\omega t} \\
        -1/2 \mathrm{e}^{-i\Delta\omega t} & -1/2 & 1/2 \mathrm{e}^{-i\Delta\omega t} & 1/2\\
        \end{pmatrix}, \\
    & A_R =
    \begin{pmatrix}
        1/2 & 1/2 \mathrm{e}^{i\Delta\omega t} & -i/2 & -i/2 \mathrm{e}^{i\Delta\omega t} \\
        1/2 \mathrm{e}^{-i\Delta\omega t} & 1/2 & -i/2 \mathrm{e}^{-i\Delta\omega t} & -i/2\\
        i/2 & i/2 \mathrm{e}^{i\Delta\omega t} & 1/2 & 1/2 \mathrm{e}^{i\Delta\omega t} \\
        i/2 \mathrm{e}^{-i\Delta\omega t} & i/2 & 1/2 \mathrm{e}^{-i\Delta\omega t} & 1/2\\
        \end{pmatrix},
    A_L =
    \begin{pmatrix}
        1/2 & 1/2 \mathrm{e}^{i\Delta\omega t} & i/2 & i/2 \mathrm{e}^{i\Delta\omega t} \\
        1/2 \mathrm{e}^{-i\Delta\omega t} & 1/2 & i/2 \mathrm{e}^{-i\Delta\omega t} & i/2\\
        -i/2 & -i/2 \mathrm{e}^{i\Delta\omega t} & 1/2 & 1/2 \mathrm{e}^{i\Delta\omega t} \\
        -i/2 \mathrm{e}^{-i\Delta\omega t} & -i/2 & 1/2 \mathrm{e}^{-i\Delta\omega t} & 1/2\\
        \end{pmatrix},
    \end{aligned}
\end{equation}
\end{center}

The expectation value for a measurement along H then equals:
\begin{equation}
    \langle A_H \rangle = Tr(\rho A_H)
\end{equation}
and similarly for the other bases, which gives: 
\begin{align}
    \braket{A_H} &= \rho_{11} + \mathrm{e}^{-i \Delta\omega t} \rho_{12} + 
        \mathrm{e}^{i \Delta\omega t} \rho_{21} + \rho_{22}  \nonumber \\ 
    \braket{A_V} &= \rho_{33} + \mathrm{e}^{-i \Delta\omega t} \rho_{34} + 
        \mathrm{e}^{i \Delta\omega t} \rho_{43} + \rho_{44}  \nonumber \\ 
    \braket{A_D} &= 
        \begin{aligned}[t]
            &1/2 (\rho_{11} + \rho_{13} + \rho_{22} + \rho_{24} + \rho_{31} + \rho_{33} + \rho_{42} +
            \rho_{44} + (\rho_{12} + \rho_{14} + \rho_{21} + \rho_{23} + \rho_{32} + \rho_{34} \\ 
            &\quad + \rho_{41} + \rho_{43}) \cos(\Delta\omega t) - i (\rho_{12} + \rho_{14} - \rho_{21} - \rho_{23} +
            \rho_{32} + \rho_{34} - \rho_{41} - \rho_{43}) \sin(\Delta\omega t))  \nonumber \\
        \end{aligned} \\
    \braket{A_A} &= 
        \begin{aligned}[t]
            &\quad 1/2 (\rho_{11} - \rho_{13} + \rho_{22} - \rho_{24} - \rho_{31} + \rho_{33} - \rho_{42} + 
            \rho_{44} + (\rho_{12} - \rho_{14} + \rho_{21} - \rho_{23} - \rho_{32} + \rho_{34} \\
            &\quad - \rho_{41} + \rho_{43}) \cos(\Delta\omega t) - i (\rho_{12} - \rho_{14} - \rho_{21} + \rho_{23} - 
            \rho_{32} + \rho_{34} + \rho_{41} - \rho_{43}) \sin(\Delta\omega t)) \nonumber \\
        \end{aligned} \\
    \braket{A_R} &= 
        \begin{aligned}[t]
            &\quad 1/2 (\rho_{11} + i \rho_{13} + \rho_{22} + i \rho_{24} - i \rho_{31} + \rho_{33} - i \rho_{42} + \rho_{44} + (\rho_{12} + i \rho_{14} + \rho_{21} + i \rho_{23} - i \rho_{32} + \rho_{34} \\
            &\quad - i \rho_{41} + \rho_{43}) \cos(\Delta\omega t) + (-i \rho_{12} + \rho_{14} + i \rho_{21} - \rho_{23} - \rho_{32} - i \rho_{34} + \rho_{41} + i \rho_{43}) \sin(\Delta\omega t)) \nonumber \\
        \end{aligned} \\
    \braket{A_L} &= 
        \begin{aligned}[t]
            &\quad 1/2 (\rho_{11} - i \rho_{13} + \rho_{22} - i \rho_{24} + i \rho_{31} + \rho_{33} + i \rho_{42} + \rho_{44} + (\rho_{12} - i \rho_{14} + \rho_{21} - i \rho_{23} + i \rho_{32} + \rho_{34} \\
            &\quad + i \rho_{41} + \rho_{43}) \cos(\Delta\omega t) + (-i \rho_{12} -\rho_{14} + i \rho_{21} + \rho_{23} + \rho_{32} - i \rho_{34} - \rho_{41} + i \rho_{43}) \sin(\Delta\omega t))  \\
        \end{aligned}\label{eqn_proj} 
\end{align}\\
We note that the expressions above contain real and imaginary parts, however assuming that $\rho$ represents a physical state (in particular, assuming a Hermitian matrix) results in real values only. By multiplying these expressions by an exponential decay over time for spontaneous emission, we obtain a model that can reproduce our experimental data for any given state. We apply a Maximum Likelihood Estimation (MLE) method, following ref.\cite{james_measurement_2001}, to fit the model to our data. This approach allows to find the most probable physical density matrix $\hat{\rho}$ accounting for the experimental data. It consists in defining $\hat{\rho}$ so that:

\begin{equation}
    \hat{\rho} = \frac{T^\dagger T}{ \text{Tr}(T^\dagger T)} \label{def_rho}
\end{equation}
where T is a triangular matrix defined as a function of 16 real parameters:
\begin{equation}
    T = \begin{pmatrix}
        t_1 & 0 & 0 & 0 \\
        t_5 + i t_6 & t_2 & 0 & 0 \\
        t_7 + i t_8 & t_9 + i t_{10} & t_3 & 0 \\
        t_{11} + i t_{12} & t_{13} + i t_{14} & t_{15} + i t_{16} & t_4 \\
    \end{pmatrix} \label{def_T}
\end{equation}
This ensures by construction that $\hat{\rho}$ fulfills the conditions of Hermiticity, positivity and unit trace.  

\begin{figure}[h]
    \centering
    \includegraphics[scale=0.9]{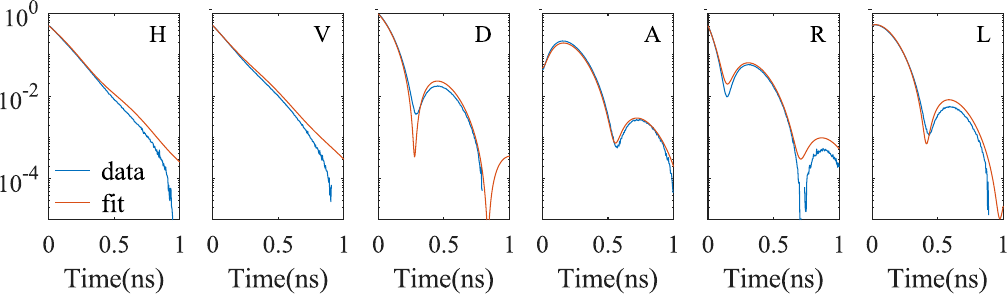}
    \caption{Fit of the polarization tomography time traces using the equations of \ref{eqn_proj}. The dataset used here corresponds to the blue triangle state in figure 3. }
    \label{figS2}
\end{figure}

The dataset used to reconstruct the density matrix corresponds to the blue triangle state in figure 3. We also implement a basis rotation of $\pi/2$ around the vertical axis of the Bloch sphere in order to set $\phi \approx 0$, removing a global phase term in the final result for clarity. The result of the fit is shown in figure \ref{figS2}. It leads to the desired density matrix $\hat{\rho}$, for which we obtain: 
\begin{equation}
    \hat{\rho} = \begin{pmatrix}
        0.4611 &	0.0084 - 0.0199i &	0.0075 - 0.0057i	& 0.4665 + 0.0102i \\
0.0084 + 0.0199i &	0.0391 &	-0.0003 + 0.0003i	& -0.0117 - 0.0022i \\
0.0075 + 0.0057i &	-0.0003 - 0.0003i &	0.0003 &	0.0080 + 0.0066i \\
0.4665 - 0.0102i &	-0.0117 + 0.0022i &	0.0080 - 0.0066i	& 0.4996  \\
    \end{pmatrix} 
\label{fittedrho}
\end{equation}

{The fidelity $F$ to a pure state $\ket{\phi}$ is then defined as:
\begin{equation}
    F(\hat{\rho}, \phi) = \braket{\phi|\hat{\rho}|\phi}.
\end{equation}
Taking $\ket{\phi}=(\ket{H,\omega_H}+\ket{V,\omega_V})/\sqrt{2}$ as the target state provides the estimated fidelity given in the main text.}

{Note that the physical density matrix $\hat{\rho}$ comprises 16 complex coefficients, which is reduced to 16 real values $t_i$ by implementing the MLE method. The time traces for the 6 polarization bases provide each an information in phase and in amplitude, yielding fewer parameters than the number of coefficients $t_i$ to fit (hence no strict unicity of solution). To obtain a fully determined system of parameters and measurements, the time traces could be completed by measuring the relative intensities resolved simultaneously in frequency and polarization, which would provide the diagonal terms $\hat{\rho}_{ii}$. However, the high contrast of the oscillations in the D/A/R/L bases provides high amplitudes for the off-diagonal terms $\hat{\rho}_{14}$ and $\hat{\rho}_{41}$ (close to the maximum value of $0.5$). In addition, the measurements in the H/V basis provide values for $\hat{\rho}_{11} + \hat{\rho}_{22} \simeq 0.5$ and $\hat{\rho}_{33} + \hat{\rho}_{44} \simeq 0.5$. These two conditions set tight bounds for the values of $\hat{\rho}_{11}$ and $\hat{\rho}_{44}$. In particular, we find that for different seed parameters, the fit systematically yields a value  $|\hat{\rho}_{14}|=|\hat{\rho}_{41}|>0.43$ with a fidelity to the target state $\ket{\phi^+}$ of $F > 86\%$. The values given in the main text correspond to the result of the fit with the most neutral seed parameters.}

\subsection{Orthogonalization and Entanglement}

{To estimate the amount of intra-particle entanglement, we can use the concurrence $C$, which is defined by 
\begin{equation}
    C(\hat{\rho}) = \lambda_1 - \lambda_2 - \lambda_3 - \lambda_4
\end{equation}
with $\lambda_{1...4}$ the eigenvalues in decreasing order of the Hermitian matrix:
\begin{equation}
    R = \sqrt{\sqrt{\hat{\rho}} \Tilde{\rho} \sqrt{\hat{\rho}}}
\end{equation}
with $\Tilde{\rho} = (\sigma_y \otimes \sigma_y) \hat{\rho}^* (\sigma_y \otimes \sigma_y)$ where $\sigma_y$ is the Pauli-y matrix. However, the entanglement concurrence is defined for a density matrix of two qubits each comprising two orthogonal states.}

{The above state tomography is performed in a natural and convenient basis, but the states $\ket{\omega_V}$ and $\ket{\omega_H}$ are pulse modes and not necessarily orthogonal and hence we cannot directly compute the concurrence of this state using the reconstructed density matrix. Instead, we first orthogonalize the frequency basis by defining a new orthogonal basis $\ket{\tilde{\omega}_H}$ and $\ket{\tilde{\omega}_V}$. We choose $\ket{\tilde{\omega}_H}=\ket{\omega_H}$ and define $\ket{\omega_V}=c_1\ket{\tilde{\omega}_H}+c_2\ket{\tilde{\omega}_V}$. By enforcing that $\braket{\tilde{\omega}_H|\tilde{\omega}_V}=0$, we have $c_1=\braket{\omega_H|\omega_V}$. Enforcing that $\ket{\omega_V}$ is normalised implies $c_2=e^{\im\phi_2}\sqrt{1-|c_1|^2}$, where we choose $\phi_2=0$ for convenience.}

{Applying this to equation (2) of the main text, we have
\begin{equation}
    \ket{\psi} = \cos(\theta)\ket{H,\tilde{\omega}_H}+\sin(\theta)\mathrm{e}^{\im\phi}\left(\frac{\im}{\im+\Delta\omega\tau}\ket{V,\tilde{\omega}_H}+\sqrt{1-\frac{1}{1+\Delta\omega^2\tau^2}}\ket{V,\tilde{\omega}_V}\right)
\end{equation}
Thus, the spectral broadening of the frequency modes puts an upper bound on the entanglement concurrence
\begin{equation}
    C \leq \sqrt{1-\frac{1}{1+\Delta\omega^2\tau^2}}.
\end{equation}}

{Although we can estimate the concurrence knowing $\Delta\omega$ and $\tau$, this assumes that the fidelity to the target state in the non-orthogonal basis is unity. To get an accurate estimate, we note that $\left|\braket{\omega_H|\omega_V}\right|^2$ is equal to the mean wavepacket overlap $M_\omega=0.318$ obtained from Hong-Ou-Mandel interference of the two frequency modes after aligning in polarization. Thus, $c_1=\sqrt{M_\omega}$. Applying the corresponding orthogonalization to the reconstructed density matrix in equation (\ref{fittedrho}) gives
\begin{equation}
    \tilde{\rho} = 
    \begin{pmatrix}
        0.4611 &	0.2670 - 0.0164i &	0.0075 - 0.0057i	& 0.38948 + 0.0052i \\
0.2670 + 0.0164i &	0.1811 &	0.0040 + 0.0030i	& 0.2215 + 0.0016i \\
0.0075 + 0.0057i &	0.0040 - 0.0030i &	0.0003 &	0.0068 + 0.0055i \\
0.38948 - 0.0052i &	0.2215 - 0.0016i &	0.0068 - 0.0055i	& 0.3483  \\
    \end{pmatrix},
\end{equation}
Finally, we can compute the concurrence of this orthogonalized density matrix to obtain the value given in the main text. Note that $M_\omega$ only provides the magnitude of $c_1$ and not the phase, which gets cancelled out in the Hong-Ou-Mandel interference measurement. However, when including an arbitrary phase, we numerically find that it has no impact on the concurrence and thus corresponds to a local phase rotation that cannot impact the amount of intra-particle entanglement.
}

\section{Model of the frequency-polarization hyperencoded qubit emission}

\subsection{Quantum dot exciton model}

{The model of the QD producing equation (1) in the main text neglects possible dissipative mechanisms that can alter the quality of the emitted state of light. To capture these imperfections, we model the emitter using a three-level system composed of the two non-degenerate excited eigenstates, $\ket{e_H}$ and $\ket{e_V}$, that decay to the QD ground state $\ket{g}$ while emitting linearly-polarized photons.}

The system Hamiltonian is $\hat{H}=\hbar\omega_H\hat{\sigma}_H^\dagger\hat{\sigma}_H+\hbar\omega_V\hat{\sigma}_V^\dagger\hat{\sigma}_V$ where $\hat{\sigma}_H=\ket{g}\!\bra{e_H}$ and $\hat{\sigma}_V=\ket{g}\!\bra{e_V}$. To simulate the dissipative dynamics of this three-level system, we use a Lindblad master equation of the form $\dot{\hat{\rho}}(t)=\mathcal{L}\hat{\rho}(t)$ with the generator $\mathcal{L}$ defined by $\mathcal{L}\hat{\rho} = -(i/\hbar)[\hat{H},\hat{\rho}]+\mathcal{L}_d\hat{\rho}$ and
\begin{equation}
    \mathcal{L}_d=\gamma_H\mathcal{D}(\hat{\sigma}_H)+\gamma_V\mathcal{D}(\hat{\sigma}_V)+\gamma_{HV}\mathcal{D}(\hat{\sigma}_H^\dagger\hat{\sigma}_V)+\gamma_{VH}\mathcal{D}(\hat{\sigma}_V^\dagger\hat{\sigma}_H)+2\gamma^\star\mathcal{D}(\ket{{g}}\!\bra{{g}})+\frac{\gamma^\star_{HV}}{2}\mathcal{D}(\hat{\sigma}_V^\dagger\hat{\sigma}_V-\hat{\sigma}_H^\dagger\hat{\sigma}_H),
\end{equation}
where $\mathcal{D}(\hat{A})\hat{\rho}=\hat{A}\hat{\rho}\hat{A}^\dagger-\{\hat{A}^\dagger\hat{A},\hat{\rho}\}/2$ is the dissipation superoperator, $\gamma_{H(V)}$ is the decay rate from the excited state $\ket{e_H}$ ($\ket{e_V}$) to the ground state $\ket{{g}}$, $\gamma_{HV(VH)}$ is the rate of population exchange from $\ket{e_H}$ to $\ket{e_V}$ ($\ket{e_V}$ to $\ket{e_H}$), $\gamma^\star$ is the pure dephasing rate between the excited state manifold and the ground state, and $\gamma^\star_{HV}$ is the pure dephasing rate of superposition states between $\ket{e_H}$ and $\ket{e_V}$. Here, we choose the convention that $\gamma^\star$ and $\gamma_{HV}^\star$ represent the coherence amplitude decay rates. That is, neglecting other dynamics, we have $\braket{e_H|\hat{\rho}|{g}}\sim e^{-\gamma^\star t-\gamma^\star_{HV} t/4}$ and $\braket{e_{H}|\hat{\rho}|e_{V}}\sim e^{-\gamma^\star_{HV} t}$. Note that $\gamma^\star$ does not degrade the coherence between states $\ket{e_H}$ and $\ket{e_V}$ but $\gamma^\star_{HV}$ will degrade the coherence between the exciton and the ground state, but at 1/4 the rate that it dephases superpositions of $\ket{e_H}$ and $\ket{e_V}$. Regardless, we expect that $\gamma^\star_{HV}\ll \gamma^\star$.

For our case, the FSS is $7.35\mu$eV and the temperature is $T=8$K, giving $k_\text{B} T \simeq 600\mu$eV. Hence, we are in the regime where $k_B T \gg \Delta_{FSS}$ and so we can simplify analytic solutions by setting $\gamma_{VH}=\gamma_{HV}$, assuming that the dissipation is primarily thermally driven. In addition, we assume that the eigenstates have approximately the same decay rates to the ground state $\gamma=\gamma_{H}=\gamma_{V}=1/\tau$. Since $\mathcal{L}$ is time independent, the dynamics of the Lindblad master equation are described by the propagation superoperator $\Lambda(t,t_0)=e^{(t-t_0)\mathcal{L}}$. The solution for any initial state $\hat{\rho}(t_0)$ of the QD is then given by $\hat{\rho}(t)=\Lambda(t,t_0)\hat{\rho}(t_0)$.

\subsection{Photonic state}

If the QD is incoherently prepared in the excited state much faster than the dissipative rates (as is the case with phonon-assisted excitation), the density matrix of the emitted photonic state will have the form
\begin{equation}
    \hat{\varrho}= p_{0}\ket{0}\!\bra{0} + p_{H}\hat{\varrho}_{HH}+ p_{V}\hat{\varrho}_{VV}+\sqrt{p_{H}p_{V}}\left(\hat{\varrho}_{HV} + \hat{\varrho}_{VH}\right)
\end{equation}
where $p_0$ is the vacuum probability and $p_{H(V)}$ is the probability of have a single photon polarized parallel to the dipole of $\ket{e_H}$ ($\ket{e_V}$). The components of the density matrix are written in terms of the orthogonal field operators $\ad_{H}$ and $\ad_{V}$ for modes with polarization parallel to the dipoles of $\ket{e_H}$ and $\ket{e_V}$, respectively. These are
\begin{equation}
\begin{aligned}
    \hat{\varrho}_{kl} &= \iint\xi_{kl}(t,t^\prime)\au_k(t)\ket{0}\!\bra{0}\ad_l(t^\prime)dtdt^\prime
\end{aligned}
\end{equation}
{where $\xi_{k,l}(t,t^\prime)$ is the joint temporal amplitude of the single photon for polarization components $k,l\in\{{H},{V}\}$. In terms of the pure state given by equation (2) of the main text, $\xi_{kl} = f_kf_l^*$ where $f_k=\sqrt{\gamma}e^{-\im\omega_k t - \gamma t/2}$ but in general $\xi_{kl}$ is not separable.}

{For a QD initially prepared in the excited state by a fast phonon-assisted excitation with excitation probability $p_\text{QD}$ and purity $\lambda$, the initial state is}
\begin{equation}
    \hat{\rho}(0) = (1-p_{QD})\ket{{g}}\!\bra{{g}}+p_{QD}\left[\cos^2(\theta)\ket{e_H}\!\bra{e_{H}}+\sin^2(\theta)\ket{e_V}\!\bra{e_{V}}+\lambda\left(\cos(\theta)\sin(\theta)\ket{e_V}\!\bra{e_{H}}e^{i\phi}+\text{h.c.}\right)\right],
\end{equation}
where $\theta$ and $\phi$ are determined by the polarization of the excitation pulse.

Making use of the input-output relations $\ad_k=\sqrt{\gamma\eta_k}\hat{\sigma}_k+\ad_{{in},k}$, and by taking the input of the collection mode to be in the vacuum state, correlation functions of the field operators that are normally-ordered and time-ordered can be evaluated from the QD master equation dynamics. {Here, $\eta_k$ is a polarization-dependent emission efficiency, which can be impacted by the dipole magnitude, cavity structure, or optical setup.} Using the quantum regression theorem, and the fact that we have at most one photon, we have
\begin{equation}
\begin{aligned}
    \xi_{kl}(t,t^\prime) &= \frac{1}{\sqrt{p_kp_l}}\braket{\au_l(t^\prime)\ad_k(t)} = \frac{\gamma\sqrt{\eta_k\eta_l}}{\sqrt{p_kp_l}}~\!\text{Tr}\!\left(\hat{\sigma}_l^\dagger \Lambda(t^\prime,t)\hat{\sigma}_k\Lambda(t,0)\hat{\rho}(0)\right)
\end{aligned}
\end{equation}
The value of $p_k$ can then be obtained by the normalization condition $\int \xi_{kk}(t,t)dt=1$.

Solving these expressions give the emission probabilities
\begin{equation}
\begin{aligned}
    p_{H} &= \frac{\eta_{H}}{2}\left[1+\left(\frac{\gamma}{\gamma+2\gamma_{HV}}\right)\cos(2\theta)\right]\\
    p_{V} &= \frac{\eta_{V}}{2}\left[1-\left(\frac{\gamma}{\gamma+2\gamma_{HV}}\right)\cos(2\theta)\right]
\end{aligned}
\end{equation}
and the temporal density functions
\begin{equation}
\label{correlationfunctionsXY}
\begin{aligned}
    \xi_{H}(t,t+{s}) &= \frac{\gamma\eta_{H}}{2p_{H}}\left(1+\cos(2\theta)e^{-2t\gamma_{HV}}\right)e^{-\gamma t - \Gamma_{s} {s}/2 + i\Delta\omega{s}/2}\\
    \xi_{V}(t,t+{s}) &= \frac{\gamma\eta_{V}}{2p_{V}}\left(1-\cos(2\theta)e^{-2t\gamma_{HV}}\right)e^{-\gamma t - \Gamma_{s} {s}/2 - i\Delta\omega{s}/2}\\
    \xi_{HV}(t,t+{s})&=\frac{\gamma\sqrt{\eta_{H}\eta_{V}}}{\sqrt{p_{H}p_{V}}}\lambda_0\cos(\theta)\sin(\theta)e^{-(\gamma+\Gamma_{p})t-\Gamma_{s}{s}/2-i\Delta\omega({s}/2+t)+i\phi},
\end{aligned}
\end{equation}
where ${s}=t^\prime-t\geq 0$, $t\geq 0$, $\Gamma_{s} = \gamma+2\gamma^\star+\gamma_{HV}+\gamma^\star_{HV}/2$ is the spectral FWHM of emission from $\ket{H}$ or $\ket{V}$, and $\Gamma_{p}=\gamma_{HV}+\gamma^\star_{HV}$ is the rate of depolarization during emission from the QD. The values for negative ${s}$ can be obtained by taking the complex conjugate of these temporal density functions, since they are Hermitian with respect to the diagonal $t=t^\prime$.

\subsection{Mean-wave packet overlap of the full state}

Measurements of the photonic state without first filtering in polarization will have direct contribution from photons in both modes $\ad_{H}$ and $\ad_{V}$ emitted from the quantum dot. The total average photon number detected is the sum of the average photons in each of the orthogonal polarization modes $\mu=\sum_k\mu_k$ where
\begin{equation}
    \mu_k=\int\braket{\au_k(t)\ad_k(t)}dt.
\end{equation}

The mean wavepacket overlap is thus given by the weighted summation of all four combinations of two photons from each mode: $\mu^2M=\sum_{k,l}\mu_k\mu_l M_{kl}$ where
\begin{equation}
    M_{kl} = \frac{1}{\mu_k\mu_l}\iint\left|\braket{\au_l(t^\prime)\ad_k(t)}\right|^2dtdt^\prime
\end{equation}
For emission containing at most one photon we have $\mu_k = p_k$ and $\braket{\au_k(t^\prime)\ad_l(t)}=\sqrt{p_kp_l}\xi_{kl}(t,t^\prime)$. Hence,
\begin{equation}
    M_{kl} = \iint\left|\xi_{kl}(t,t^\prime)\right|^2dtdt^\prime,
\end{equation}
is the trace purity of the $\hat{\varrho}_{kl}$ component of the density matrix. Then, the total unfiltered mean wavepacket overlap is given by
\begin{equation}
    M = \frac{p_{H}^2M_{HH}+p_{H}p_{V}\left(M_{HV}+M_{VH}\right)+p_{V}^2M_{VV}}{(p_{H}+p_{V})^2}
\end{equation}
Evaluating $M_{kl}$ using Eq. (\ref{correlationfunctionsXY}), we have
\begin{equation}
    \begin{aligned}
        M_{HH} &= \frac{\gamma^2\eta_{H}^2}{4p_{H}^2\Gamma_\text{s}}\left(\frac{1}{\gamma}+\frac{2\cos(2\theta)}{\gamma+\gamma_{HV}}+\frac{\cos^2(2\theta)}{\gamma+2\gamma_{HV}}\right)\\
        M_{HV}&=M_{VH}=\frac{\gamma^2\eta_{H}\eta_{V}\lambda_0^2\cos^2(\theta)\sin^2(\theta)}{p_{H}p_{V}\Gamma_\text{s}(\gamma+\Gamma_\text{p})}\\
        M_{VV} &= \frac{\gamma^2\eta_{V}^2}{4p_{V}^2\Gamma_\text{s}}\left(\frac{1}{\gamma}-\frac{2\cos(2\theta)}{\gamma+\gamma_{HV}}+\frac{\cos^2(2\theta)}{\gamma+2\gamma_{HV}}\right)\\
    \end{aligned}
\end{equation}
Note that all of the components of $M$ are independent of $\Delta_{FSS}$. This is because $\Delta_{FSS}$ gives rise to a time-dynamic frequency/polarization evolution that is perfectly canceled during Hong-Ou-Mandel interference with another wavepacket that has the same frequency/polarization evolution. However, they do depend on the excitation polarization angle $\theta$ in a fairly non-trivial way as shown in figure 5.b. of the main.

{Finally, we also compute the overlap between the two spectral modes:
\begin{equation}
\begin{aligned}
    M_\omega &= \iint \xi_H(t,t^\prime)\xi^*_V(t,t^\prime)dtdt^\prime\\
    &=\frac{\gamma\eta_H\eta_V\left(\gamma+4\gamma_{HV}-\gamma\cos(4\theta)\right)}{8p_Hp_V\left(\gamma+2\gamma_{HV}\right)\left(\Gamma_s-\im\Delta\omega\right)},
\end{aligned}
\end{equation}
which reduces to the ideal pure state solution $M_\omega=\im/(\im+\Delta\omega\tau)$ when $\theta=\pi/4$, $\gamma_{HV}=0$, and $\Gamma_s=\gamma$.}

\bibliography{Biblio-DF-NC}